\documentstyle[amsfonts,preprint,aps]{revtex}

\begin{document}
\draft
\title{{\bf Global Superdiffusion of Weak Chaos}}
\author{{\bf Itzhack Dana}}
\address{Minerva Center and Department of Physics, Bar-Ilan University, 
Ramat-Gan 52900, Israel}
\maketitle

\begin{abstract}
A class of kicked rotors is introduced, exhibiting accelerator-mode
islands (AIs) and {\em global} superdiffusion for {\em arbitrarily weak}
chaos. The corresponding standard maps are shown to be exactly related
to generalized web maps taken modulo an ``oblique cylinder''. Then, in a
case that the web-map orbit structure is periodic in the phase plane,
the AIs are essentially {\em normal} web islands folded back into the
cylinder. As a consequence, chaotic orbits sticking around the AI
boundary are accelerated {\em only} when they traverse tiny {\em
``acceleration spots''}. This leads to chaotic flights having a
quasiregular {\em steplike} structure. The global weak-chaos
superdiffusion is thus basically different in nature from the
strong-chaos one in the usual standard and web maps.\newline
\end{abstract}


The complexity and rich variety of dynamical behaviors of Hamiltonian
systems with two degrees of freedom is well exhibited by simple 1D
time-periodic models. A realistic paradigm is the kicked rotor with
Hamiltonian $H=p^{2}/2+V(x)\sum_{n=-\infty }^{\infty }\delta (t-n)$, where 
$p$ is angular momentum, $x$ is angle on a circle ${\cal S}=[-\pi ,\ \pi )$,
and the potential $V(x)$ is usually chosen as $V(x)=\kappa \cos x$, $\kappa $
being a nonintegrability parameter. The phase space is a cylinder and the
values of $(x,\ p)$ at integer times $t=n-0$ are related by the ``standard''
map \cite{bvc,jmg} 
\begin{equation}
{\bf \Phi }_{{\rm s}}:\ \ p_{n+1}=p_{n}+f(x_{n}),\ \ \ \
x_{n+1}=x_{n}+p_{n+1}\ \ \text{mod }{\cal S},  \label{sm}
\end{equation}
where $f(x)=-dV/dx=\kappa \sin x$. For $\kappa >\kappa _{{\rm c}}\approx
0.9716$, this map features {\em global} chaos, i.e., a connected chaotic
region unbounded in the $p$ direction and ranging from $x=-\pi $ to $x=\pi $ 
\cite{jmg}. The average kinetic energy of an ensemble in this region grows
diffusively, $\left\langle p_{n}^{2}/2\right\rangle \propto n$ \cite{bvc}.
For $\kappa $ significantly larger than $\kappa _{{\rm c}}$ (strong-chaos
regime), there emerge accelerator-mode islands (AIs) \cite{ai,kz,sd} moving
ballistically ($p_{n}\propto n$). Then, chaotic orbits sticking around the
boundary of an AI perform also ballistic motion. This leads to long chaotic 
{\em ``flights''} and to {\em superdiffusion} of the global chaos,
$\left\langle p_{n}^{2}/2\right\rangle \propto n^{\mu }$ ($1<\mu <2$)
\cite{ai,sd}.\newline

In this paper, we introduce a class of kicked rotors exhibiting AIs and
{\em global} superdiffusion for {\em arbitrarily weak} chaos. These
systems correspond to standard maps (\ref{sm}) with
\begin{equation}
f(x)=Ks(x)+\kappa g(x),  \label{fx}
\end{equation}
where $-4<K<-2$, $\kappa $ is a perturbation parameter, $s(x)$ is the
sawtooth function [$s(x)=x$ for $x\in {\cal S}$ and $s(x+2\pi )=s(x)$], and
$g(x)$ is a general smooth $2\pi $-periodic function. We denote these maps,
for $-4<K<0$, by ${\bf \Phi }_{{\rm s}}^{(K,\kappa )}$. For $\kappa =0$,
${\bf \Phi }_{{\rm s}}^{(K,0)}$ exhibits generically a pseudochaotic behavior
(zero Lyapunov exponent) \cite{pa}; important exceptions, with regular
elliptic motion, are the cases of integer $K=-3,\ -2,\ -1$ \cite{cm}. For
small $\kappa $, weak chaos emerges, see Fig. 1. The superdiffusion for
$-4<K<-2$ is usually characterized by large values of $\mu $ (for $\kappa =0$,
$\mu \approx \mu _{{\rm \max }}=2$). To get a better understanding of the
nature of the AIs and the chaotic flights responsible for the
superdiffusion, we first show that ${\bf \Phi }_{{\rm s}}^{(K,\kappa )}$ is
exactly related to a ``web map'' \cite{sw,da} taken modulo an ``oblique
cylinder''. A generalized web map ${\bf \Phi }_{{\rm w}}$ on a phase plane
${\bf w}\equiv (u,\ v)$ (column vector) is defined by: 
\begin{equation}
{\bf \Phi }_{{\rm w}}:\ {\bf w}_{n+1}=A\cdot \lbrack {\bf w}_{n}+{\bf F}
({\bf w}_{n})],  \label{wm}
\end{equation}
where $A=(\cos \alpha ,\ \sin \alpha ;\ -\sin \alpha ,\ \cos \alpha )$ is
the matrix for a rotation by angle $\alpha $ and ${\bf F}({\bf w})$ is a
vector function periodic in ${\bf w}$. For integer $K$, the orbit structure
of ${\bf \Phi }_{{\rm w}}$ is periodic in the phase plane. Then, the AIs of
${\bf \Phi }_{{\rm s}}^{(K,\kappa )}$ for $K=-3$ are essentially {\em normal}
(nonaccelerating) islands of ${\bf \Phi }_{{\rm w}}$ folded back into the
cylinder. This fact makes these AIs basically different from usual AIs \cite
{kz} and has a significant impact on the nature of the chaotic flights, as
we show by studying in detail the case of $K=-3$ with $g(x)=\sin x$. While
elliptic orbits deep inside an AI move ballistically, chaotic orbits
sticking to the AI boundary perform ballistic motion {\em only} when they
pass through very small regions which we call {\em ``acceleration spots''}.
The rest (the majority) of the AI boundary behaves like that of a normal
island, so that the chaotic orbits perform a completely {\em bounded} motion
on it. This gives rise to chaotic flights having a quasiregular {\em steplike}
structure which becomes increasingly pronounced as $\kappa \rightarrow 0$.
\newline

To establish the relation between ${\bf \Phi }_{{\rm s}}^{(K,\kappa )}$ and
${\bf \Phi }_{{\rm w}}$, consider first a map ${\bf \Psi }^{(K,\kappa )}$
obtained from (\ref{sm}) by removing the mod ${\cal S}$\ and by using 
(\ref{fx}) with $s(x)$ replaced by $x$. This map, defined on the entire phase
plane of ${\bf z}\equiv (x,\ p)$, can be written as follows: 
\begin{equation}
{\bf \Psi }^{(K,\kappa )}:\ {\bf z}_{n+1}=B\cdot \lbrack {\bf z}_{n}+
{\bf G}({\bf z}_{n})],  \label{lsm}
\end{equation}
where $B$ is the matrix $(K+1,\ 1;\ K,\ 1)$ and ${\bf G}({\bf z})=\kappa
g(x)(0,\ 1)$. For $-4<K<0$, one can easily verify that $B=Q^{-1}AQ$, where
$A=(\cos \alpha ,\ \sin \alpha ;\ -\sin \alpha ,\ \cos \alpha )$ with 
\begin{equation}
2\cos \alpha =K+2  \label{aK}
\end{equation}
and $Q=(k,\ -k/2;\ 0,\ 1/2)$ with $k\equiv \tan (\alpha /2)$. Thus, the
composition $Q\circ {\bf \Psi }^{(K,\kappa )}\circ Q^{-1}$ is precisely a
web map (\ref{wm}) with ${\bf w}=(u,\ v)=Q\cdot {\bf z}$ and ${\bf F}({\bf w})
=Q\cdot {\bf G}(Q^{-1}\cdot {\bf w})$. Explicitly, ${\bf F}({\bf w})
=(\kappa /2)g(v+u/k)(-k,\ 1)$. We denote by ${\cal C}=[-\pi ,\ \pi )\times
(-\infty ,\ \infty )$ the cylindrical phase space for ${\bf \Phi }_{{\rm s}}
^{(K,\kappa )}$\ and define ${\bf \Psi }_{{\cal C}}^{(K,\kappa )}$ as the
map (\ref{lsm}) with $x_{n+1}$ (in first equation) taken modulo ${\cal S}$.
Clearly, if we restrict ourselves to initial conditions ${\bf z}_{0}\in 
{\cal C}$, ${\bf \Psi }_{{\cal C}}^{(K,\kappa )}={\bf \Phi }_{{\rm s}}
^{(K,\kappa )}$. The image of ${\cal C}$ under $Q$ is an ``oblique
cylinder'' ${\cal C}_{\alpha }$, i.e., the strip bounded by the lines $v=\pm
\pi -u/k$. We define the web map ``modulo ${\cal C}_{\alpha }$'' as 
\begin{equation}
\widetilde{{\bf \Phi }}_{{\rm w}}={\bf \Phi }_{{\rm w}}\ {\rm mod}\ {\cal C}
_{\alpha }:\ {\bf w}_{n+1}=A\cdot \lbrack {\bf w}_{n}+{\bf F}({\bf w}_{n})]
-m\widetilde{{\bf w}},  \label{wmm}
\end{equation}
where $\widetilde{{\bf w}}=Q\cdot (2\pi ,\ 0)=(2\pi k,\ 0)$ and $m$ is the
unique integer such that ${\bf w}_{n+1}\in $ ${\cal C}_{\alpha }$. The
following exact relation then holds for all orbits with initial conditions
${\bf z}_{0}\in {\cal C}$ (or ${\bf w}_{0}\in {\cal C}_{\alpha }$): 
\begin{equation}
{\bf \Phi }_{{\rm s}}^{(K,\kappa )}={\bf \Psi }_{{\cal C}}^{(K,\kappa
)}=Q^{-1}\circ \widetilde{{\bf \Phi }}_{{\rm w}}\circ Q\text{.}  \label{sw}
\end{equation}

For integer $K=-3,\ -2,\ -1$ [corresponding, by Rel. (\ref{aK}), to $q\equiv
2\pi /\alpha =3,\ 4,\ 6$, respectively], it is easy to show that the orbit
structure of ${\bf \Psi }^{(K,\kappa )}$ is periodic with unit cell ${\Bbb T}
^{2}=[-\pi ,\ \pi )^{2}$. This implies a similar periodicity for ${\bf \Phi }
_{{\rm w}}=Q\circ {\bf \Psi }^{(K,\kappa )}\circ Q^{-1}$ with unit cell
${\Bbb T}_{\alpha }^{2}=Q\cdot {\Bbb T}^{2}$. In fact, ${\bf \Phi }_{{\rm w}}$
has crystalline symmetry (triangular, square, hexagonal for $q=3,\ 4,\ 6$,
respectively) \cite{sw,da}. One can then expect the existence of an extended
chaotic orbit having this symmetry and forming a ``stochastic web''. This
web, which has been observed for particular maps ${\bf \Phi }_{{\rm w}}$ 
\cite{sw,da}, encircles the torus ${\Bbb T}_{\alpha }^{2}$ in two
independent directions. This implies {\em global} chaos for the map 
(\ref{wmm}) in ${\cal C}_{\alpha }$ and, due to (\ref{sw}), also for 
${\bf \Phi }_{{\rm s}}^{(K,\kappa )}$ in ${\cal C}$.\newline

For noninteger $K$, the maps ${\bf \Psi }^{(K,\kappa )}$ and ${\bf \Phi }_
{{\rm w}}$ exhibit no periodicity and there is no simple relation between the
orbit structures of ${\bf \Phi }_{{\rm s}}^{(K,\kappa )}$ and ${\bf \Phi }_
{{\rm w}}$. In the $\kappa =0$ case, theoretical arguments and numerical
evidence \cite{pa} strongly indicate that for irrational $q$ the torus
${\Bbb T}^{2}$ can be partitioned into two regions having nonzero area: (a) 
A connected pseudochaotic region (zero Lyapunov exponent). (b) An apparently
dense set of elliptic islands whose boundaries do not cross or touch the
discontinuity line $x=-\pi $. Because of the last fact, the pseudochaotic
region encircles ${\Bbb T}^{2}$ in both the $x$ and $p$ directions, implying
{\em global} pseudochaos in ${\cal C}$. This will generally turn into global 
weak chaos when a small perturbation $\kappa g(x)$ is applied. Fig. 1(b)
shows an example for $\alpha =\pi (\sqrt{5}-1)/2$ ($K\approx -2.7247$) and 
$\kappa =0.15$.\newline

Accelerator-mode fixed points (AFPs) of (\ref{sm}) satisfy $p_{1}-p_{0}=2\pi
j$ and $x_{1}=x_{0}=x_{1}-p_{1}+2\pi j^{\prime }$ for integers $j\neq 0$ and 
$j^{\prime }$. One can choose $p_{0}=0$ and, for ${\bf \Phi }_{{\rm s}}
^{(K,0)}$ ($\kappa =0$), we get $x_{0}=2\pi j/K$. Since $x_{0}\in (-\pi ,\
\pi )$, one must have $j=\pm 1$ and $-4<K<-2$. The latter results remain
essentially unchanged for sufficiently small $\kappa $. The AFPs are
surrounded by relatively large AIs (see Fig. 1) and a strong superdiffusion
(large $\mu $) was always observed numerically for $-4<K<-2$. In what
follows, we shall study in detail the case of $K=-3$ with $g(x)=\sin x$ on
the basis of Rel. (\ref{sw}). Fig. 2 shows the stochastic web of ${\bf \Phi }
_{{\rm w}}$ ($q=3$) for $\kappa =0.8$. The ``cylinder'' ${\cal C}_{\alpha }$
is the oblique strip bounded by the parallel dashed lines. Together with
these lines, the two horizontal dashed segments define the unit cell ${\Bbb T
}_{\alpha }^{2}$ in which there appears, up to the transformation $Q$ in
(\ref{sw}), the chaotic region in Fig. 1(a). The $j$th unit cell in ${\cal C}
_{\alpha }$, $j=-\infty ,...,\ \infty $, contains one ``hexagonal'' island 
H$_{j}$ and two ``triangular'' islands, L$_{j}$ and R$_{j}$. The hyperbolic
(x) points $a$, $b$, $d$, and $e$ lie on the boundaries of ${\cal C}_{\alpha
}$ while $c$ and $f$ are inside ${\cal C}_{\alpha }$. The points $a$, $b$, 
$c$ are equivalent, modulo ${\Bbb T}_{\alpha }^{2}$, to $d$, $e$, $f$,
respectively. It is interesting to see first how the AFPs emerge according
to (\ref{sw}). The islands H$_{j}$, L$_{j}$, R$_{j}$ are invariant under
${\bf \Phi }_{{\rm w}}^{3}$ and their centers CH$_{j}$, CL$_{j}$, CR$_{j}$
are fixed points of ${\bf \Phi }_{{\rm w}}^{3}$ which are rotated clockwise
by $\alpha =2\pi /3$ under ${\bf \Phi }_{{\rm w}}$. Then, denoting ${\bf
\Phi }_{{\rm w}}$ by $\mapsto$ and ``modulo ${\cal C}_{\alpha }$'' by
$\Rightarrow$, it is clear from Fig. 2 that CL$_{-1}\mapsto$ CL$_{0}\mapsto$
CL$_{0}^{\prime }\Rightarrow $ CL$_{1}\mapsto$ CL$_{1}^{\prime }\Rightarrow$
CL$_{2}$. In general, we see that $\widetilde{{\bf \Phi }}_{{\rm w}}( 
$CH$_{j})=$ CH$_{j}$, $\widetilde{{\bf \Phi }}_{{\rm w}}($CL$_{j})=$
CL$_{j+1}$, and $\widetilde{{\bf \Phi }}_{{\rm w}}($CR$_{j})=$
CR$_{j-1}$, so that CL$_{j}$ and CR$_{j}$ correspond to the AFPs and L$_{j}$ 
and R$_{j}$ correspond to the AIs.\newline

Extensive numerical observations indicate that for small $\kappa $ a chaotic
orbit is always a random sequence of three types of motion: (a) A
``H-motion'', bounded in $p$, sticking around the boundary of H$_{j}$. (b) A
``flight'' in the positive or (c) negative $p$ direction, accompanied by
stickiness to the boundary of L$_{j}$ or R$_{j}$, respectively. Remarkably,
a flight was always found to be a quasiregular {\em sequence of steps}. An
example of such a flight, interrupted by a H-motion, is shown in Fig. 3 for
$\kappa =0.3$. Very long, uninterrupted flights (of at least $10^{6}$
iterations) were usually observed for $\kappa \leq 0.3$. The steps in the
flights clearly leave their fingerprints in the strong superdiffusion of
$\langle p_{n}^{2}/2\rangle $, at least for small $n$ (see inset in Fig. 3).
This steplike structure will now be explained using Rel. (\ref{sw}). The
notation $a($L$_{j})$ (and similar notation for other x points and islands)
will indicate a web (chaotic) point sticking to the L$_{j}$ boundary very
close to $a$ and {\em inside} ${\cal C}_{\alpha }$; if this point is {\em
outside} ${\cal C}_{\alpha }$, it will be denoted by $\overline{a}($L$_{j})$.
For small $\kappa >0$, ${\bf \Phi }_{{\rm w}}$ is almost a clockwise
rotation by $\alpha =2\pi /3$ while the motion of web points under ${\bf
\Phi }_{{\rm w}}^{3}$ is a slow drift in the directions of the stable and
unstable manifolds, indicated by arrows in Fig. 2. Since the drift velocity
vanishes near x points, a cycle such as $e($L$_{-1})\mapsto f($L$_{0})\mapsto 
d($L$_{0}^{\prime })$ can repeat many times. This cycle will
then drift to the cycle $c($L$_{-1})\mapsto a($L$_{0})\mapsto b($L$_{0}^
{\prime })$ which can also repeat many times. The cycles $efd$ and $cab$
give the ``horizontal'' part of a step, where $p$ is bounded around its
values at the x points (see Fig. 4). This part contains $l$ cycles ($l=66$
in Fig. 4), where $l$ is the largest integer such that all points ${\bf \Phi 
}_{{\rm w}}^{i}[e($L$_{-1})]$, $i=0,...,\ 3l-1$, lie {\em inside} ${\cal C}
_{\alpha }$, so that no modulo ${\cal C}_{\alpha }$ has to be taken in the
cycles. The ``vertical'' part of a step is due to the following process. The
last $cab$ cycle is followed by the cycle $c($L$_{-1})\mapsto a($L$_{0})
\mapsto \overline{b}($L$_{0}^{\prime })$, where $\overline{b}($L$_{0}^
{\prime })$ is {\em outside} ${\cal C}_{\alpha }$. Taking then the
modulo ${\cal C}_{\alpha }$, $\overline{b}($L$_{0}^{\prime })\Rightarrow 
e($L$_{1})$, we see that $a($L$_{0})$ is mapped into $e($L$_{1})$ by 
$\widetilde{{\bf \Phi }}_{{\rm w}}$. Next, $\widetilde{{\bf \Phi }}_{{\rm w}}
[e($L$_{1})]=f($L$_{2})$. Thus, since $c$ is equivalent to $f$, the cycle 
$cae$ is actually an ``acceleration'' cycle $fae$ with $\widetilde{{\bf
\Phi }}_{{\rm w}}^{3}[f($L$_{-1})]=f($L$_{2})$. The set $\{a($L$_{0})\}$ of 
all points $a($L$_{0})$ which are mapped into points $\overline{b}($L$_{0}^
{\prime })$ by ${\bf \Phi }_{{\rm w}}$ is the {\em ``acceleration spot''} 
(AS) in unit cell $j=0$. As shown by the inset in Fig. 4, the AS touches the 
line $x=-\pi $ ($x^{\prime }=0$), corresponding to the lower boundary of 
${\cal C}_{\alpha }$. The cycle $fae$ will repeat $r$ times ($r=6$ in Fig. 4), 
where $r$ is the largest integer such that all points $\widetilde{{\bf 
\Phi }}_{{\rm w}}[f($L$_{3i-1})]\ \mathop{\rm mod}\ {\Bbb T}_{\alpha }^{2}$, 
$i=0,...,\ r-1$, lie in the AS. After $r$ $fae$ cycles, one leaves the AS
by crossing the line $x=-\pi $ and arrives to $\overline{a}($L$_{0})$ which is
equivalent to $d($L$_{0}^{\prime })$. Thus, $\widetilde{{\bf \Phi }}_
{{\rm w}}[f($L$_{3r-1})]$ is equivalent to $d($L$_{0}^{\prime })$, {\em not} 
to $a($L$_{0})$. The horizontal $efd$ cycles of the next step then start.
This completes the analysis of one step.\newline

Fig. 4 shows a strong trapping near six islands in the AS. In fact, by
considering a large number of steps in very long orbits, we found that $r$
assumes only two values for $\kappa =0.3$: $r=6$ with probability $P\approx
0.95$ and $r=7$ with $P\approx 0.05$. The value of $l$ ranges between $l=53$
and $l=82$. This range should be associated with the continuation of the
trapping, outside the AS, near the island chain to which the six AS islands
belong. As $\kappa $ decreases, the steplike structure of the flights
becomes more pronounced and the high regularity of the vertical parts of the
steps, i.e., the values of $r$, continues to be observed. For $\kappa =0.2$, 
$l=88-136$ and $r=11$ ($P\approx 0.99$) or $r=12$ ($P\approx 0.01$). For
$\kappa =0.1$, $l=150-259$ and $r=29$ ($P\approx 0.93$) or $r=30$ ($P\approx
0.07$).\newline

In conclusion, our study of the $K=-3$ case indicates that the global
superdiffusion of weak chaos is basically different in nature from the
superdiffusion observed in the usual standard and web maps \cite{ai,sd}.
In the latter systems, the AIs are {\em ``tangle''} islands \cite{kz}.
These islands born in a strong-chaos regime and are fundamentally
different from normal islands, e.g., resonance or web islands, which
continue to exist in an integrable limit. Since a tangle island lies
inside the lobe of a turnstile \cite{kz}, it causes the acceleration of
chaotic orbits sticking {\em all around} its boundary. On the other
hand, Rel. (\ref{sw}) implies that the AIs for $K=-3$ are essentially
normal web islands folded back into the cylinder. As a consequence, one
can have a situation that a chaotic orbit sticking to the AI boundary is
accelerated {\em only} at tiny ``acceleration spots'', in sharp contrast
with the case of tangle islands. The resulting steplike structure of the
chaotic flights is gradually assumed also by elliptic flights with
initial conditions approaching the AI boundary from inside the AI. The
basic origin of both the AIs and the acceleration spots is the
folding-back mechanism. Quantum manifestations of the superdiffusion in
the usual standard map are well known \cite{sz,adl} and have been
observed in experimental realizations of the quantum kicked rotor
\cite{er}. The perturbed sawtooth map (\ref{sm}) with (\ref{fx})
corresponds to a kicked rotor with a nonsmooth potential. The quantum
version of such a system is experimentally realizable by, e.g., an
optical analogue \cite{jk}. It should be then interesting to study the
fingerprints of the new kind of superdiffusion in the corresponding
quantum systems. Quite recently \cite {qc1,qc2}, the quantum sawtooth
map was found to be a suitable model for quantum computation of complex
dynamics. The exponential-decay rate of the concurrence (measure of
quantum entanglement) was shown to be proportional to the classical
diffusion coefficient \cite{qc2}. The extension of this study to the
perturbed sawtooth map, with its new kind of chaotic-transport
properties, thus appears to be a natural and interesting future
task.\newline

This work was partially supported by the Israel Science Foundation
administered by the Israel Academy of Sciences and Humanities.

\figure{Fig. 1. Global chaos for the standard map (\ref{sm}) with
(\ref{fx}), $g(x)=\sin x$, and: (a) $K=-3$, $\kappa =0.8$; (b)
$K=-2.7247498$, $\kappa =0.15$. The AIs are the regions indicated by
elliptic orbits.}

\figure{Fig. 2. Stochastic web of the web map ${\bf \Phi }_{{\rm w}}$
related by (\ref{sw}) to the standard map defined in the caption of Fig.
1(a). See text for details.}

\figure{Fig. 3. Chaotic flight, interrupted by a H-motion, for $K=-3$,
$\kappa =0.3$, and $g(x)=\sin x$. The inset shows $\ln (\langle
E_n\rangle)$ versus $\ln (n)$ (dotted line), where $E_n=p_n^2/2$, the
average $\langle\ \rangle$ is over an ensemble of $10^6$ initial
conditions well localized around $(x=0 ,\ p=-\pi )$, and $n_{\rm
max}=12000$; the linear fit (solid line) has slope $\mu \approx 1.96$.}

\figure{Fig. 4. Magnification of one step in the chaotic flight shown in
Fig. 3. The ``horizontal" part of the step (dots) consists of $efd$
cycles followed by $cab$ cycles (total of $198$ points). The ``vertical"
part (squares) consists of six ``acceleration"  cycles $fae$ ($18$
points). The inset shows the ``acceleration spot" $\{a($L$_{0})\}$ in
unit cell $j=0$ using the variables $x^{\prime }=(x+\pi )\cdot 10^5$ and
$p^{\prime }=(p+\pi )\cdot 10^5$.}

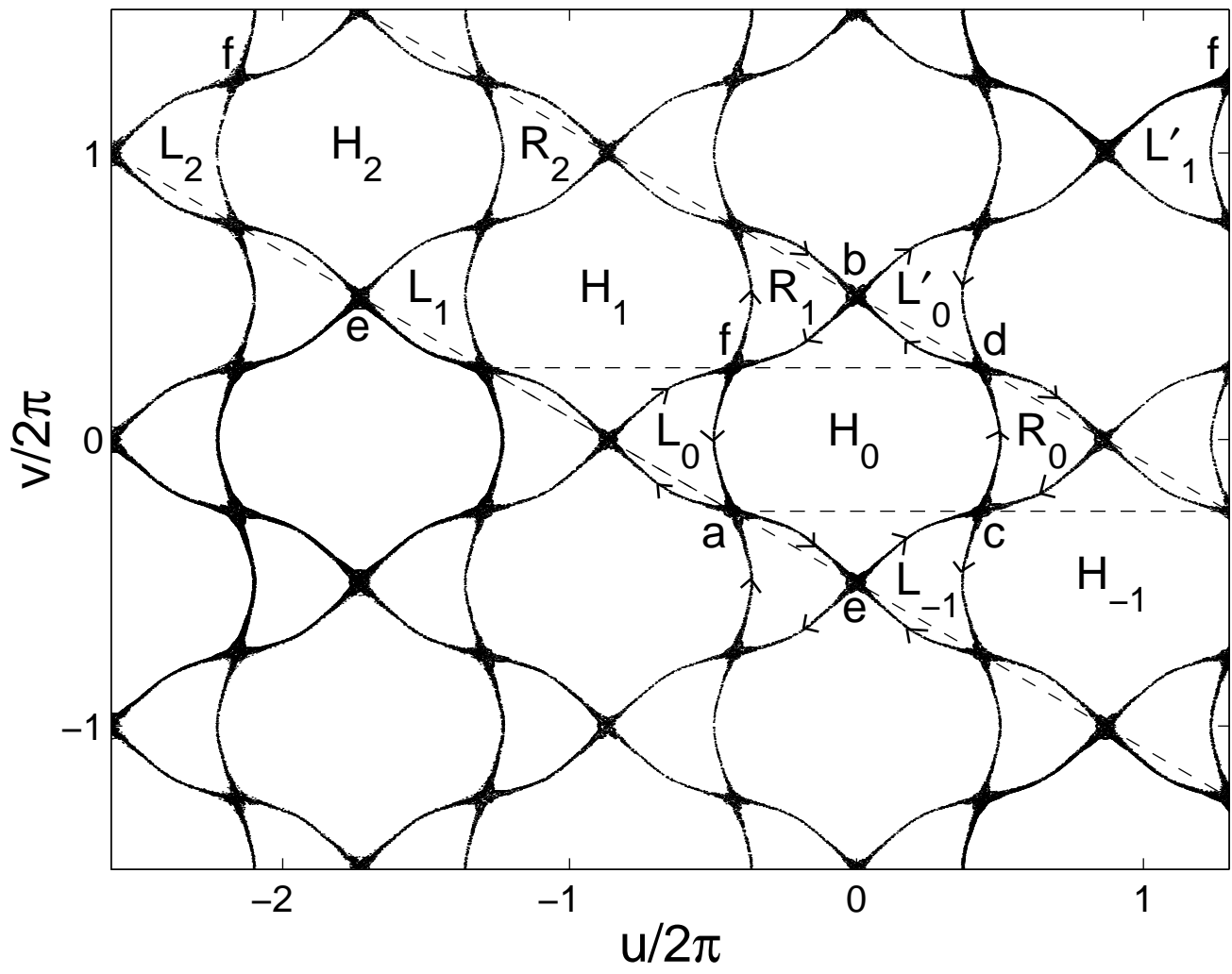
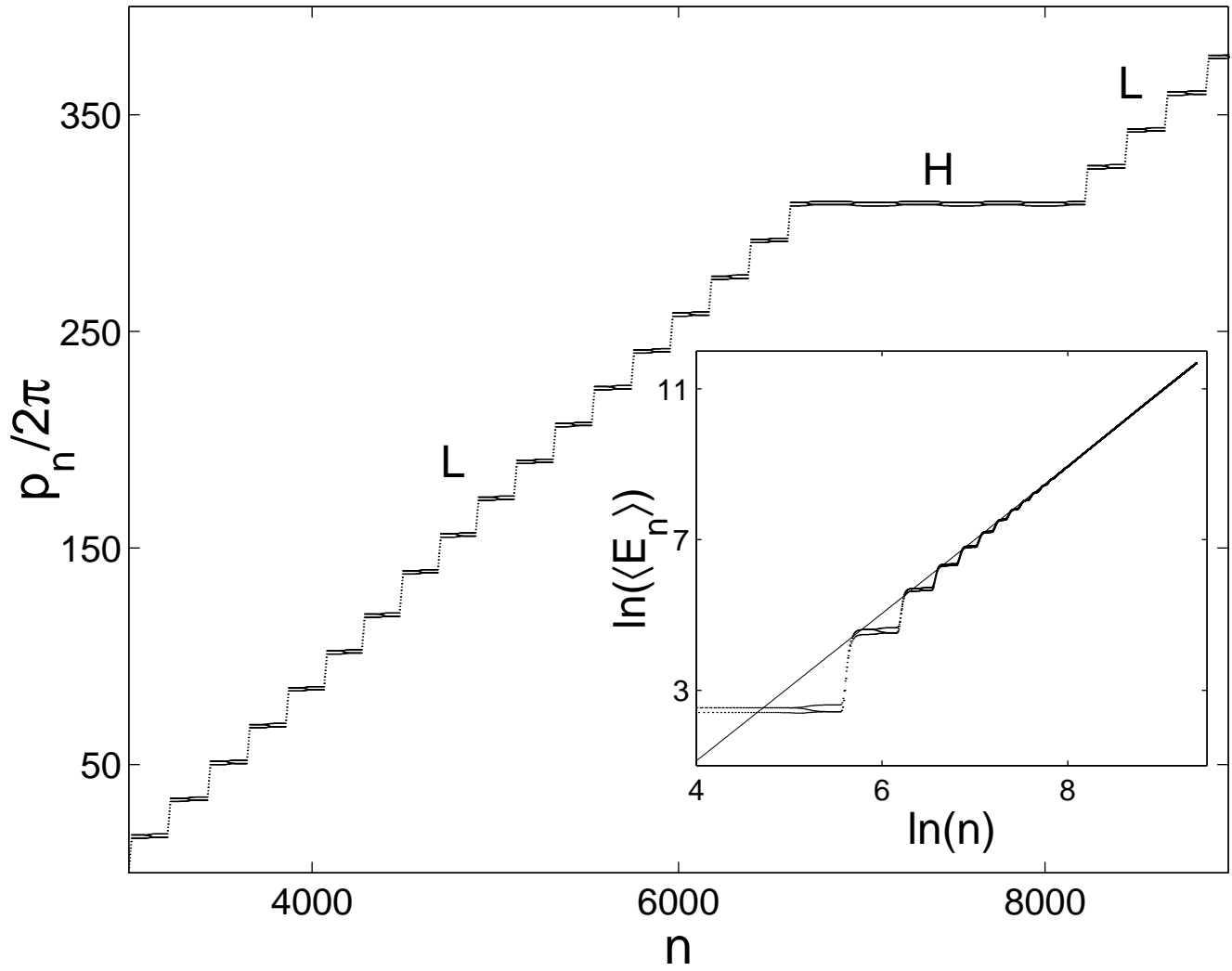
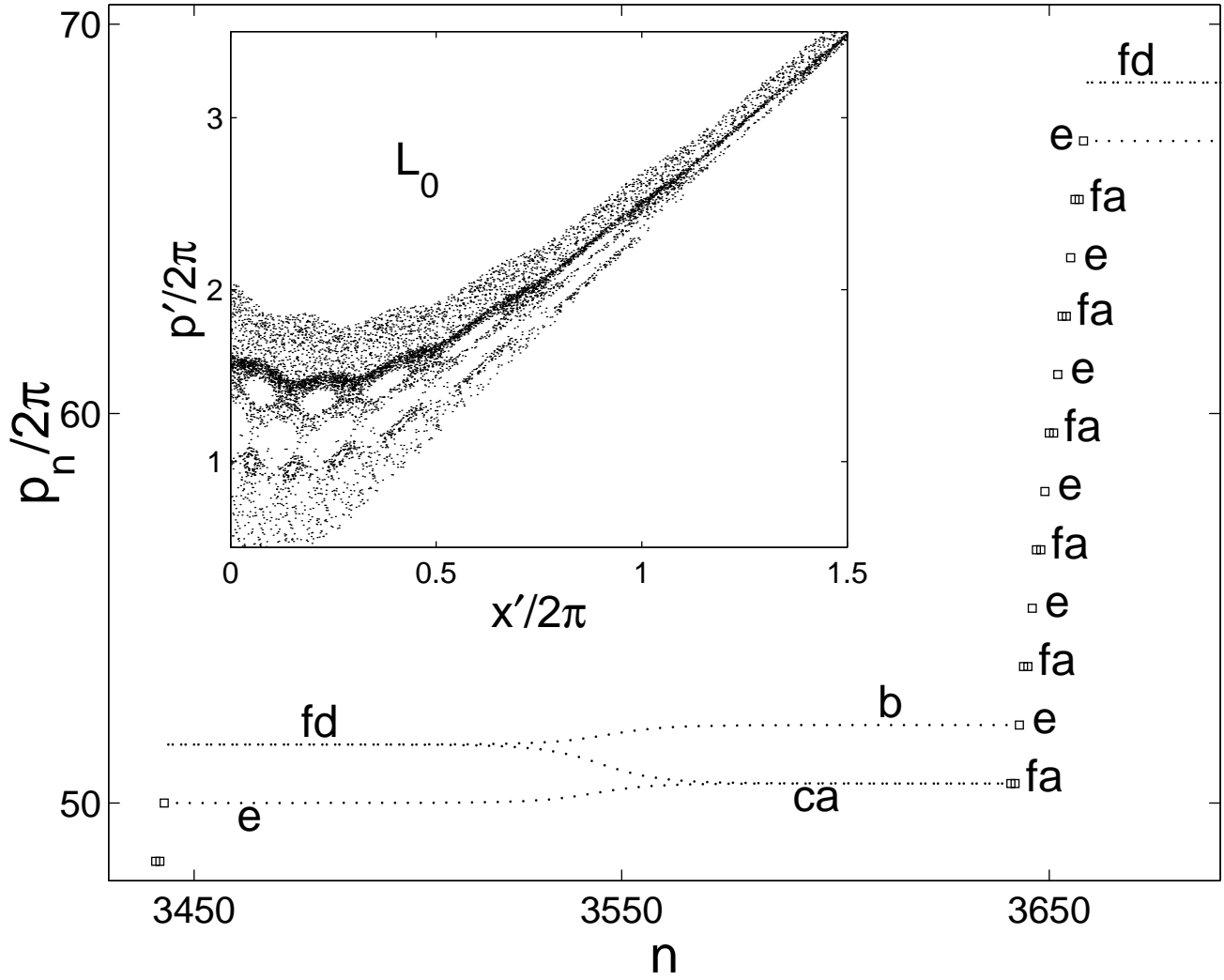
\end{document}